\documentclass[5p,times]{elsarticle}
\usepackage{xspace}
\usepackage{graphicx}
\usepackage[cmex10]{amsmath}
\usepackage{amssymb}
\usepackage{algorithm}
\usepackage{algorithmicx}
\usepackage[noend]{algpseudocode}
\usepackage{xcolor}
\usepackage{url}
\usepackage{booktabs}
\usepackage{subcaption}
\usepackage{flushend}
\usepackage{lineno,hyperref}

\newcommand\newtilde{\raise.17ex\hbox{$\scriptstyle\sim$}}

\algdef{SE}[DOWHILE]{Do}{doWhile}{\algorithmicdo}[1]{\algorithmicwhile\ #1}%
\algnotext{EndFor}
\algnotext{EndIf}
\algnotext{EndWhile}
\algnotext{EndProcedure}

\journal{Parallel Computing}

\begin{document}
\begin{frontmatter}

\title{Parallel Graph Coloring Algorithms for Distributed GPU Environments}

\author{Ian Bogle and George M. Slota}
\address{Department of Computer Science}
\address{Rensselaer Polytechnic Institute}
\address{Lally 317, 110 8th St}
\address{Troy, NY 12180}
\address{\{boglei,slotag\}@rpi.edu}

\author{Erik G. Boman, Karen D. Devine, and Sivasankaran Rajamanickam}
\address{Scalable Algorithms Department}
\address{Sandia National Laboratories, New Mexico}
\address{\{egboman,kddevin,srajama\}@sandia.gov}

\maketitle

\begin{abstract}
Graph coloring is often used in parallelizing scientific computations that run in distributed and multi-GPU environments; it identifies sets of independent data that can be updated in parallel. 
Many algorithms exist for graph coloring on a single GPU or in distributed memory, but to the best of our knowledge, hybrid MPI+GPU algorithms have been unexplored until this work.
We present several MPI+GPU coloring approaches based on the distributed coloring algorithms of 
Gebremedhin et al. 
and the shared-memory algorithms of 
Deveci et al. 
The on-node parallel coloring uses implementations in 
KokkosKernels, 
 which provide parallelization for both multicore CPUs and GPUs.
We further extend our approaches to compute distance-2 and partial distance-2
colorings, giving the first known distributed, multi-GPU algorithm for these problems.
In addition, we propose a novel heuristic to reduce communication for recoloring in distributed graph coloring. 
Our experiments show that our approaches operate efficiently on inputs too large to fit on a single GPU and scale up to graphs with 76.7 billion edges running on 128 GPUs.



\end{abstract}


\end{frontmatter}



\section{Introduction}

We present new multi-GPU, distributed memory implementations of distance-1,
distance-2, and partial distance-2 graph coloring.
\emph{Distance-1 graph coloring} assigns \emph{colors} (i.e., labels) to all vertices in a graph such that no two neighboring vertices have the same color. 
Similarly, \emph{distance-2 coloring} assigns colors such that no vertices within \emph{two hops}, also called a ``two-hop neighborhood,'' have the same color. 
\emph{Partial distance-2 coloring} is a special case of distance-2 coloring, in which only one set of a bipartite graph's vertices are colored.
Usually, these problems are formulated as NP-hard optimization problems, where the number of colors used to fully color a graph is minimized.
Serial heuristic algorithms have traditionally been used to solve these problems, one of the most notable being the DSatur algorithm of Br{\'e}laz~\cite{brelaz1979new}.
More recently, parallel algorithms~\cite{IAB:deveci2016parallel,IAB:bozdaug2008framework} have been proposed; such algorithms usually require multiple \emph{rounds} to correct for improper \emph{speculative} colorings produced in multi-threaded or distributed environments.

There are many useful applications of graph coloring.
Most commonly, it is employed to find concurrency in parallel scientific computations~\cite{IAB:deveci2016parallel, IAB:allwright1995comparison}; all data sharing a color can be updated in parallel without incurring race conditions.
Other applications use coloring as a preprocessing step to speed up the computation of Jacobian and Hessian matrices~\cite{IAB:gebremedhin2013colpack} and to identify short circuits in printed circuit designs~\cite{IAB:garey1976application}.
Despite the intractability of minimizing the number of colors for non-trivial graphs, such applications benefit from good heuristic algorithms that produce small numbers of colors.
For instance, Deveci et al.~\cite{IAB:deveci2016parallel} show that a smaller number of colors used by a coloring-based preconditioner reduces the runtime of a conjugate gradient solver by 33\%.


In particular, this work is motivated by the use of graph coloring as a preprocessing step for distributed scientific computations such as automatic differentiation~\cite{IAB:gebremedhin2020introduction}. 
For such applications, assembling the associated graphs on a single node to run a sequential coloring algorithm may not be feasible~\cite{IAB:bozdaug2008framework}.
As such, we focus on running our algorithms on the parallel architectures used by the underlying applications.
These architectures typically are highly distributed, with multiple CPUs and/or GPUs per node.
Therefore, we specifically consider coloring algorithms that can use the ``MPI+X'' paradigm, where the Message Passing Interface (MPI) library is used in distributed memory and ``X'' is multicore CPU or GPU acceleration.


\subsection{Contributions}

We present and examine two MPI+X implementations of distance-1 coloring as well as one MPI+X implementation of distance-2 coloring.
In order to run on a wide variety of architectures, we use the Kokkos performance portability framework~\cite{IAB:edwards2014kokkos,kokkoskernels} for on-node parallelism and Trilinos~\cite{IAB:heroux2005overview} for distributed MPI-based parallelism.
The combination of Kokkos and MPI allows our algorithms to run on multiple multicore CPUs or multiple GPUs in a system.
For this paper, we focus on the performance of our algorithms in MPI+GPU environments.
For distance-1 coloring of real-world networks, our algorithms achieve up to 2.38x speedup on 128 GPUs compared to a single GPU, and only a 2.23\% increase in the number of colors on average.
For distance-2 coloring, our algorithm achieves up to 33x speedup and, on average, a 7.5\% increase in the number of colors.
We also demonstrate good weak scaling behavior up to 128 GPUs for graphs with up to 12.8 billion vertices and 76.7 billion edges.

\section{Background}

\subsection{Coloring Problem }

While there exist many definitions of the ``graph coloring problem,'' we specifically consider variants of distance-1 and distance-2 coloring.
Consider graph $G = (V,E)$ with vertex set $V$ and edge set $E$.
\emph{Distance-1 coloring} assigns to each vertex $v \in V$ a color $C(v)$ such that $\forall (u,v) \in E, C(u) \neq C(v)$.
In \emph{distance-2 coloring}, colors are assigned such that 
$\forall (u,v),(v,w) \in E, C(u) \neq C(v) \neq C(w)$; 
i.e., all vertices within two hops of each other have different colors.
\emph{Partial Distance-2 coloring} is a special case of distance-2 coloring in which
$\forall (u,v),(v,w) \in E, C(u) \neq C(w)$; it is typically applied to bipartite graphs
in which only one set of the vertices is given colors (thus, the designation ``partial'').
Partial distance-2 coloring is used to color sparse Jacobian matrices~\cite{GebremedhinMannePothen}.
When a coloring satisfies one of the above constraints, it is called \emph{proper}.
The goal is to find proper colorings of $G$ such that the total number of different colors used is minimized.




\subsection{Coloring Background}

While minimizing the number of colors is NP-hard, serial coloring algorithms using greedy heuristics have been effective for many applications~\cite{IAB:gebremedhin2000scalable}.
The serial greedy algorithm in Algorithm~\ref{IAB:alg:serialgreed} colors vertices one at a time.
Colors are represented by integers, and the smallest usable color is assigned as a vertex's color.
Most serial and parallel coloring algorithms use some variation of greedy coloring, with algorithmic differences usually involving the processing order of vertices or, in parallel, the handling of conflicts and communication.

\begin{algorithm}
  \caption{Serial greedy coloring algorithm}
  \label{IAB:alg:serialgreed}
  \begin{algorithmic}
    \Procedure{SerialGreedy}{Graph $G=(V,E)$}
      \State $C(\forall v \in V) \gets 0$ \Comment{Initialize all colors as null}
      \ForAll{$v \in V$ in some order}
        \State $c \gets$ the \emph{smallest} color not used by a neighbor of $v$
        \State $C(v) \gets c$
      \EndFor
    \EndProcedure
  \end{algorithmic}
\end{algorithm}

\emph{Conflicts} in a coloring are edges that violate the color-assignment criterion; for example, in distance-1 coloring, a conflict is an edge with both endpoints sharing the same color.
Colorings that contain conflicts are not proper colorings, and are referred to as \emph{pseudo-colorings}.
Pseudo-colorings arise only in parallel coloring, as conflicts arise only when two vertices are colored concurrently.
A coloring's ``quality'' refers to the number of colors used; higher quality colorings of a graph $G$ use fewer colors, while lower quality colorings of $G$ use more colors.


It has been observed that the order vertices are visited affects the number of colors needed.
Popular vertex orderings include largest-degree-first, smallest-degree-last, and saturation degree~\cite{Besta20}. 
These orderings are highly sequential and do not allow much parallelism. However, relaxations of those orderings
can allow some parallelism \cite{Hasenplaugh14}.


\subsection{Parallel Coloring Algorithms}

There are two popular approaches to parallel graph coloring.
The first concurrently finds independent sets of vertices and concurrently colors all of the vertices in each set. This approach was used by 
Jones and Plassmann~\cite{IAB:jones1993parallel}.
Osama et al.~\cite{osama19} found independent sets on a single GPU and explored the impact of varying the baseline independent set algorithm.

The second approach, referred to as ``speculate and iterate''~\cite{IAB:gebremedhin2000scalable,IAB:ccatalyurek2012graph}, colors as many vertices as possible in parallel and then iteratively fixes conflicts in the resulting pseudo-coloring until no conflicts remain.
Gebremedhin et al.~\cite{IAB:gebremedhin2000scalable}, {\c{C}}ataly{\"{u}}rek et al.~\cite{IAB:ccatalyurek2012graph} and Rokos et al.~\cite{IAB:rokos2015fast} present shared-memory implementations based on the speculate and iterate approach.
Deveci et al.~\cite{IAB:deveci2016parallel} present implementations based on the speculate and iterate approach that are scalable on a single GPU.
Distributed-memory algorithms such as those in~\cite{IAB:bozdaug2008framework,IAB:sariyuce2012scalable} use the speculate and iterate approach. 
Grosset et al.~\cite{IAB:grosset2011evaluating} present a hybrid speculate and iterate approach that splits computations between the CPU and a single GPU, 
but does not operate on multiple GPUs in a distributed memory context.
Sallinen et al.~\cite{Sallinen16} demonstrated how to color very large, dynamic graphs efficiently.
Besta et al. \cite{Besta20} developed shared memory coloring algorithms and analyzed their performance.
They compared to both Jones-Plassman and speculative methods, but only on multicore CPU.

Bozda{\u{g}} et al.~\cite{IAB:bozdaug2008framework} showed that, in distributed memory, the speculative approach is more scalable than methods based on the independent set approach of Jones and Plassmann.
Therefore, we choose a speculative and iterative approach with our algorithms.

\subsection{Distributed Coloring}


In a typical distributed memory setting, an input graph is split into subgraphs that are assigned to separate processes.
A process's \emph{local graph} $G_l = \{V_l+V_g, E_l+E_g\}$ is the subgraph assigned to the process.
Its vertex set $V_l$ contains \emph{local vertices}, and a process is said to \emph{own} its local vertices.  The intersection of all processes' $V_l$ 
is null, and the union equals $V$.
The local graph also has non-local vertex set $V_g$, with such non-local vertices commonly referred to as \emph{ghost vertices}; these vertices are copies of 
vertices owned by other processes.
To ensure a proper coloring, each process needs to store color state information for both local vertices and ghost vertices; typically, ghost vertices are treated as read-only.
The local graph contains edge set $E_l$, edges between local vertices, and $E_g$, edges containing at least one ghost vertex as an endpoint.
Bozda{\u{g}} et al.~\cite{IAB:bozdaug2008framework} also defines two subsets of local vertices: \emph{boundary vertices} and \emph{interior vertices}.
Boundary vertices are locally owned vertices that share an edge with at least one ghost; interior vertices are locally owned vertices that do not neighbor ghosts.
For processes to communicate colors associated with their local vertices, each vertex has a unique global identifier (GID).
\section{Methods}

We present three hybrid MPI+GPU algorithms, called Distance-1 (D1), Distance-1 Two Ghost Layer (D1-2GL) and Distance-2 (D2). D1 and D1-2GL solve the distance-1 coloring problem, and D2 does distance-2 coloring.
We apply a variation of our D2 coloring to do partial D2-coloring (PD2).
We leverage Trilinos~\cite{IAB:heroux2005overview} for distributed MPI-based 
parallelism and Kokkos~\cite{IAB:edwards2014kokkos} for on-node parallelism. KokkosKernels~\cite{kokkoskernels} provides baseline implementations of distance-1 and distance-2 coloring algorithms that we use and modify for our local coloring and recoloring subroutines.

Our three proposed algorithms follow the same basic framework, which builds upon that of Bozda{\u{g}} et al.~\cite{IAB:bozdaug2008framework}.
Bozda{\u{g}} et al. observe that interior vertices can be properly colored independently on each process without creating conflicts or requiring communication.
They propose first coloring interior vertices, and then coloring boundary vertices in small batches over multiple rounds involving communication between processes. 
This approach can reduce the occurrence of conflicts, which in turn reduces the amount of communication necessary to properly color the boundary. 
In our approach, we color all \emph{local} vertices first.
Then, after communicating boundary vertices' 
colors, we fix all conflicts.  Several
rounds of conflict resolution and communication may be needed to resolve all
conflicts.
We found that this approach was generally faster than the batched boundary
coloring, and it allowed us to use existing parallel coloring routines in KokkosKernels without substantial modification.

\begin{algorithm}[!htb]
  \caption{Distributed-Memory Speculative Coloring}
  \label{IAB:alg:overview}
  \begin{algorithmic}
    \Procedure{Parallel-Color} {\newline \hspace*{0.5pc} Local Graph $G_l=\{V_l+V_g,E_l+E_g\}$,GID}
      \State colors $\gets$ Color($G_l$, colors) \Comment{Initially color local graph}
      \State Communicate colors of boundary vertices
      \State conflicts $\gets$ Detect-Conflicts($G_l$, colors, GID)
      \State Allreduce(conflicts, SUM) \Comment{Global sum conflicts}
      \While{conflicts $>$ 0}
	\State $\mathit{gc} \gets$ current colors of all ghosts
	\State colors = Color($G_l$, colors) \Comment{Recolor conflicted}
        \State \Comment{vertices}
	\State Replace ghost colors with $\mathit{gc}$
        \State Communicate updated boundary colors
        \State conflicts $\gets$ Detect-Conflicts($G_l$, colors, GID)
        \State Allreduce(conflicts, SUM) \Comment{Global sum conflicts}
      \EndWhile
      \State \textbf{return} colors
    \EndProcedure
  \end{algorithmic}
\end{algorithm}

Algorithm~\ref{IAB:alg:overview} demonstrates the general approach for our three speculative distributed algorithms. 
First, each process colors all local vertices with a shared-memory algorithm.
Then, each process communicates its boundary vertices' colors to processes with corresponding ghosts.
Processes detect conflicts in a globally consistent way and remove the colors of conflicted vertices.
Finally, processes locally recolor all uncolored vertices, communicate updates, detect conflicts, and repeat until no conflicts are found.

\subsection{Distributed Boundaries}

\begin{figure}
  \includegraphics[scale=0.2]{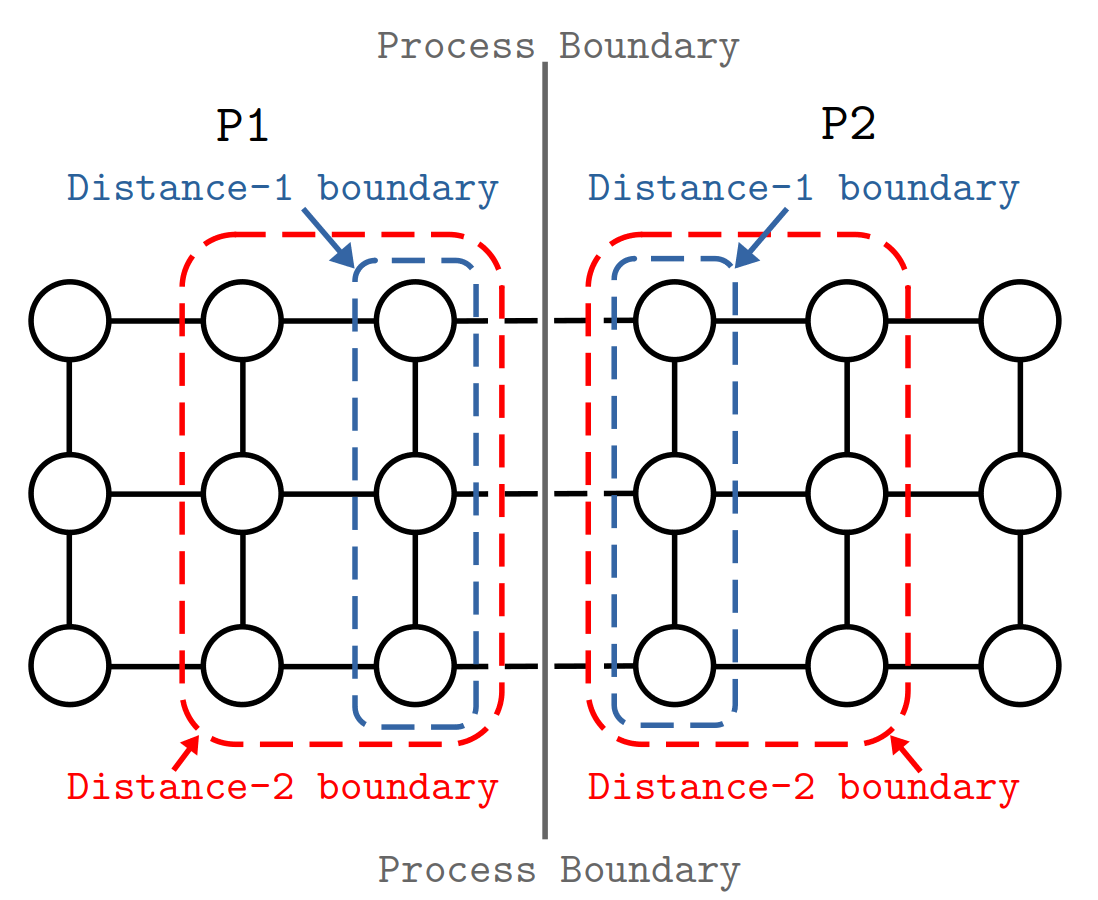}
  \caption{Definition of boundary vertex sets for different coloring instances}
  \label{IAB:boundary-verts}
\end{figure}

Figure~\ref{IAB:boundary-verts} shows the sets of boundary vertices for distance-1 and distance-2 formulations of graph coloring.
A process' distance-1 boundary vertices are its owned vertices that have edge neighbors owned by other processes.
Its distance-2 boundary vertices are its owned vertices whose neighbors have neighbors owned by other processes.
These sets allow us to optimize our distributed conflict detection, as only vertices in the boundary may conflict with a vertex on another process.

\subsection{Distance-1 Coloring (D1)}

Our Distance-1 method begins by independently coloring all owned vertices on each process using the GPU-enabled algorithms by Deveci et al.~\cite{IAB:deveci2016parallel}
VB\_BIT and EB\_BIT in KokkosKernels~\cite{kokkoskernels}.
VB\_BIT uses vertex-based parallelism; each vertex is colored by a single thread. VB\_BIT uses compact bit-based representations of colors to make it performant on GPUs.
EB\_BIT uses edge-based parallelism; a thread colors the endpoints of a single edge. EB\_BIT also uses the compact color representation to reduce memory usage on GPUs.
For graphs with skewed degree distribution (e.g., social networks), edge-based parallelism typically yields better workload balance between GPU threads.
We observed that for graphs with a sufficiently large maximum degree, edge-based EB\_BIT outperformed vertex-based VB\_BIT on Tesla V100 GPUs.
Therefore, we use a simple heuristic based on maximum degree:  we use EB\_BIT for graphs with maximum degree greater than 6000; otherwise, we use VB\_BIT.

\begin{algorithm}[!htb]
\algrenewcommand\algorithmicindent{1.0em}
  \caption{Distance-1 conflict detection}
  \label{IAB:alg:conflictres}
  \begin{algorithmic}
    \Procedure{Detect-Conflicts-D1}{\newline \hspace*{0.5pc} Local Graph $G_l=\{V_l+V_g,E_l+E_g\}$, colors, GID} 
      \State conflicts $\gets$ 0
      \ForAll{$v \in V_g$} \textbf{in parallel}
        \ForAll{$\langle v, u\rangle \in (E_g)$}
          \State conflicts $\gets$ conflicts $+$ Check-Conflicts($v,u,\ldots$)
	  \If{colors[$v$] $=$ 0}
	    \State \textbf{break}
	  \EndIf
        \EndFor
      \EndFor
      \State \textbf{return} conflicts
    \EndProcedure
  \end{algorithmic}
\end{algorithm}

\begin{algorithm}[!htb]
  \caption{Algorithm to identify and resolve conflicts}
  \label{IAB:alg:conflictcheck}
  \begin{algorithmic}
    \Procedure{Check-Conflicts}{$v$, $u$, colors, GID, recolorDegrees}
      \If{colors[$v$] $=$ colors[$u$]}
	\If{recolorDegrees and degree($v$) $<$ degree($u$)}
	  \State colors[$v$] $\gets$ 0
	\ElsIf{recolorDegrees and degree($u$) $<$ degree($v$)}
	  \State colors[$u$] $\gets$ 0
        \ElsIf{rand(GID[$v$]) $>$ rand(GID[$u$])}
	  \State colors[$v$] $\gets$ 0
	\ElsIf{rand(GID[$u$]) $>$ rand(GID[$v$])}
	  \State colors[$u$] $\gets$ 0
	\ElsIf{GID[$v$] $>$ GID[$u$]}
	  \State colors[$v$] $\gets$ 0
	\Else
	  \State colors[$u$] $\gets$ 0
	\EndIf
        \State \textbf{return} 1
      \EndIf
      \State \textbf{return} 0
    \EndProcedure
  \end{algorithmic}
\end{algorithm}

Algorithm~\ref{IAB:alg:conflictres} shows the conflict detection component of Algorithm~\ref{IAB:alg:overview}. 
This algorithm runs on each process using its local graph $G_l$.
It detects conflicts across processor boundaries and uncolors vertices to 
resolve the conflicts before recoloring.

After the initial coloring, only boundary vertices can be in conflict with one another\footnote{As suggested by Bozda{\u{g}} et al., we considered reordering local vertices to group all boundary vertices together for ease of processing. This optimization did not show benefit in our implementation, as reordering tended to be slower than coloring of the entire local graph.}.
We perform a full exchange of boundary vertices' colors using collective communication functions implemented in the Zoltan2 package of Trilinos~\cite{IAB:heroux2005overview}.
After the initial all-to-all boundary exchange, we only communicate the colors of boundary vertices that have been recolored.
After each process receives its ghosts' colors, it detects conflicts by checking each owned vertex's color against the colors of its neighbor.
The conflict detection is done in parallel over the owned vertices using Kokkos.
The overall time of conflict detection is small enough that any imbalance resulting from our use of vertex-based parallelism is insignificant relative to end-to-end times for the D1 algorithm.


Once we have identified all conflicts, we again use VB\_BIT or EB\_BIT to recolor the determined set of conflicting vertices. 
We modified KokkosKernels' coloring implementations to accept a ``partial'' coloring and the full local graph, including ghosts. 
(Our initial coloring phase did not need ghost information.)
We also modified VB\_BIT to accept a list of vertices to be recolored. 
Such a modification was not feasible for EB\_BIT.

Before we detect conflicts and recolor vertices, we save a copy of the ghosts' colors ($\mathit{gc}$ in Algorithm~\ref{IAB:alg:conflictres}).
Then we give color zero to all vertices that will be recolored; our coloring functions interpret color zero as uncolored.

To prevent the coloring functions from resolving conflicts without respecting our conflict resolution rules (thus preventing convergence of our parallel coloring), we allow a process to temporarily recolor some ghosts,
even though the process does not have enough color information to correctly recolor them.  The ghosts' colors are then 
restored to their original values in order to keep ghosts' colors consistent with their owning process.
Then, we communicate only recolored owned vertices, ensuring that recoloring changes only owned vertices.

\subsection{Distributed Recoloring Using Vertex Degrees}

When a conflict is found, only one vertex involved in the conflict needs to be recolored. 
Since conflicts happen on edges between two processes' vertices, both processes must agree on which vertex will be recolored.

We propose a new algorithm for selecting vertices to be recolored in the conflict phase, based on prioritizing by vertex degrees. 
This idea was inspired by the effectiveness of largest-first and smallest-last ordering in the serial greedy algorithm. 
To the best of our knowledge, prioritizing the distributed recoloring of lower degree vertices is a novel approach to distributed coloring conflict resolution.
In this approach, 
shown in Algorithm~\ref{IAB:alg:conflictcheck}, when recolorDegrees is true, our conflict detection prioritizes recoloring the lower degree vertex involved in a distributed conflict.
For vertices with equal degree, we adopt the random conflict resolution scheme of Bozda{\u{g}} et al. 
in which 
the conflicted vertex with the higher random number generated from its global identifier (GID) is chosen for recoloring.

The idea behind our recolorDegrees heuristic is that recoloring vertices with large degrees will likely result in giving those vertices a higher color, while recoloring vertices with a smaller degree may be able to use a smaller color for that vertex. 
Additionally, recoloring vertices with fewer neighbors means that it is less likely that we recolor neighboring vertices concurrently which can reduce the number of conflicts that arise during distributed recoloring.
We show that this approach generally decreases runtime for distance-1 coloring, and reduces the number of colors used.
In our experiments, recolorDegrees reduces our color usage by 8.9\% and runtime by roughly 7\% for D1 on average. 
It achieves a maximum speedup of 45\%, and a maximum color reduction of 39\% over using D1 without recolorDegrees.

We compute the vertex degrees only once. Possible variations include using a ``dynamic'' 
degree based on how many neighbors have been colored or the ``saturation degree'' (how many colors the colored neighbors have been assigned). We do not investigate those variations here.

\subsection{Two Ghost Layers Coloring (D1-2GL)}

Our second algorithm for distance-1 coloring, D1-2GL, follows the D1 method, but adds another ghost vertex ``layer'' to the subgraphs on each process.
In D1, a process' subgraph does not include neighbors of ghost vertices unless those neighbors are already owned by the process.
In D1-2GL, we include all neighbors of ghost vertices (the two-hop neighborhood of local vertices) in each process's subgraph, giving us ``two ghost layers.''
To the best of our knowledge, this approach has not been explored before with respect to graph coloring.

This method can reduce the total amount of communication relative to D1 for certain graphs by reducing the total number of recoloring rounds needed.
In particular, for mesh or otherwise regular graphs, the second ghost layer is primarily made up of interior vertices on other processes.
Interior vertices are never recolored, so the colors of the vertices in the second ghost layer are fixed. Each process can then directly resolve more conflicts in a consistent way, thus requiring fewer rounds of recoloring.
Fewer recoloring rounds results in fewer collective communications.

However, in D1-2GL, each communication is more expensive than in D1, because a larger boundary from each process is communicated.
Also, in irregular graphs, the second ghost layer often does not have mostly interior vertices.
The relative proportion of interior vertices in the second layer also gets smaller as the number of processes increases.
For the extra ghost layer to pay off, it must reduce the number of rounds of communications enough to make up for the increased cost of each communication. 

To construct the second ghost layer on each process, processes exchange the adjacency lists of their boundary vertices; this step is needed only once.
After the ghosts' connectivity information is added, we use the same coloring approach as in D1.

We optimize our conflict detection for both distance-1 implementations by looking through only the ghost vertices' adjacencies ($E_g$), as they neighbor all local boundary vertices.
Our local coloring algorithms require our local graphs to have undirected edges to ghost vertices, so this optimization is trivial for both D1 and D1-2GL.

\begin{table*}[!t]
  \centering
  \caption{Summary of D1 and D2 input graphs. $\delta_{avg}$ refers to average degree and $\delta_{max}$ refers to maximum degree. Values listed are after preprocessing to remove multi-edges and self-loops. k = thousand, M = million, B = billion.}
  \begin{tabular}{|r|r|r|r|r|r|r|}
    \hline
    Graph   & Class   & \#Vertices  & \#Edges  & $\delta_{avg}$ & $\delta_{max}$ & Memory (GB)\\
    \hline
    ldoor           & PDE Problem     & 0.9 M   & 21 M    & 45  & 77  & 0.32 \\
    Audikw\_1       & PDE Problem     & 0.9 M   & 39 M    & 81  & 345 & 0.59 \\
    Bump\_2911      & PDE Problem     & 2.9 M   & 63 M    & 43  & 194 & 0.96 \\
    Queen\_4147     & PDE Problem     & 4.1 M   & 163 M   & 78  & 89  & 2.5 \\
    soc-LiveJournal1& Social Network  & 4.8 M   & 43 M    & 18  & 20 k & 0.67 \\
    hollywood-2009  & Social Network  & 1.1 M   & 57 M    & 99  & 12 k & 0.86 \\
    twitter7        & Social Network  & 42 M    & 1.4 B   & 35  & 2.9 M & 21 \\
    com-Friendster  & Social Network  & 66 M    & 1.8 B   & 55  & 5.2 k & 27 \\
    europe\_osm     & Road Network    & 51 M    & 54 M    & 2.1 & 13    & 1.2 \\
    indochina-2004  & Web Graph       & 7.4 M   & 194 M   & 26  & 256 k & 2.9 \\
    MOLIERE\_2016 & Document Mining Network & 30 M & 3.3 B & 80 & 2.1 M & 49 \\
    rgg\_n\_2\_24\_s0 & Synthetic Graph & 17 M    & 133 M   & 15    & 40 & 2.1\\
    kron\_g500-logn21 & Synthetic Graph & 2.0 M   & 182 M   & 87    & 8.7 & 2.7\\
    mycielskian19     & Synthetic Graph & 393 k   & 452 M   & 2.3 k & 196 k & 6.7\\
    mycielskian20     & Synthetic Graph & 786 k   & 1.4 B   & 3.4 k & 393 k & 21\\
    \hline
    hexahedral & Weak Scaling Tests & 12.5 M -- 12.8 B & 75 M -- 76.7 B & 6 & 6 & 1.2 GB -- 1.1 TB\\
    \hline
  \end{tabular}\\
  \label{IAB:tab:graphs}
\end{table*}

\subsection{Distance-2 Coloring (D2)}

Our distance-2 coloring algorithm, D2, builds upon both D1 and D1-2GL.
As with distance-1 coloring, we use algorithms from Deveci et al. in KokkosKernels for local distance-2 coloring.
Specifically, we use NB\_BIT, which is a ``net-based'' distance-2 coloring algorithm that uses the approach described by Ta{\c{s}} et al.~\cite{IAB:tacs2017greed}. 
Instead of checking for distance-2 conflicts only between a single vertex and its two-hop neighborhood, the net-based approach detects distance-2 conflicts among the immediate neighbors of a vertex.
Our D2 approach also utilizes a second ghost layer to give each process the full two-hop neighborhood of its boundary vertices.
This enables each process to directly check for distance-2 conflicts with local adjacency information. 
To find a distance-2 conflict for a given vertex, its entire two-hop neighborhood must be checked for potential conflicting colors.

\begin{algorithm}
\algrenewcommand\algorithmicindent{1.0em}
\caption{Distance-2 conflict detection}
\label{IAB:alg:d2con}
\begin{algorithmic}
\Procedure{Detect-D2-Conflicts}{\newline \hspace*{0.5pc} Local Graph $G_l=\{V_l+V_g,E_l+E_g\}$, $V_b$, colors, GID, doPartialColoring} 
  \State conflicts $\gets$ 0
  \ForAll{$v \in V_b$} \textbf{in parallel}
    \ForAll{$\langle v, u\rangle \in (E_l+E_g)$} 
      \If{not doPartialColoring}
        \State conflicts $\gets$ conflicts $+$ Check-Conflicts($v,u,\ldots$)
        \If{colors[$v$] $=$ 0}
	  \State \textbf{break}
        \EndIf
      \EndIf
      \ForAll{$\langle u, x\rangle \in (E_l+E_g)$} 
	\State \Comment{$u$ is one hop and $x$ is two hops from $v$}
	\State conflicts $\gets$ conflicts $+$ Check-Conflicts($v,x,\ldots$)
	\If{colors[$v$] $=$ 0}
	  \State \textbf{break}
	\EndIf
      \EndFor
      \If{colors[$v$] $=$ 0}
	\State \textbf{break}
      \EndIf
    \EndFor
  \EndFor
\State \textbf{return} conflicts
\EndProcedure
\end{algorithmic}
\end{algorithm}

Algorithm~\ref{IAB:alg:d2con} shows conflict detection in D2 for each process.
We again use vertex-based parallelism while detecting conflicts; each thread examines the entire two-hop neighborhood of a vertex $v$.
The input argument $V_b$ is the set of distance-2 boundary vertices (as in Figure~\ref{IAB:boundary-verts}), which we precompute.
As with distance-1 conflict detection, we identify all local conflicts and use a random number generator to ensure that vertices to be recolored are chosen consistently across processes.
The iterative recoloring method of D1 then also works for D2 --- we recolor all conflicts, replace the old ghost colors, and then communicate local changes.

\subsection{Partial Distance-2 Coloring (PD2)}\label{IAB:method:PD2}

We have also implemented an algorithm, PD2, that solves the partial distance-2 coloring problem.
Partial distance-2 coloring is similar to distance-2 coloring, but it detects and resolves only two-hop conflicts.
Typically, partial distance-2 coloring is used on non-symmetric graphs.  A bipartite graph 
$B(V_s, V_t, E_B)$ is constructed from $G(V,E)$ with an undirected edge 
$\langle v_s \in V_s, v_t \in V_t \rangle$ $\in E_B$ 
for each directed edge $\langle v_s, v_t \rangle \in E$; colors are needed only for vertices in $V_s$.
Partial distance-2 coloring colors only one set of the vertices in the bipartite graph, 
which is why it is a partial coloring.
In algorithm~\ref{IAB:alg:d2con}, when doPartialColoring is false, the algorithm detects all distance-2 conflicts. 
When doPartialColoring is true, it only detects two-hop conflicts for the partial coloring.
Currently, our PD2 implementation must color all vertices in the bipartite representation of the graph;
applications can ignore colors for vertices in $V_t$.
Removing this limitation is a subject for future work.

\subsection{Partitioning}

We assume that target applications partition and distribute their input graphs in some way before calling these coloring algorithms. In our experiments, we used XtraPuLP v0.3~\cite{slota2017partitioning} to partition our graphs. 
Determining optimal partitions for coloring is not our goal in this work.
Rather, we have chosen a partitioning strategy representative of that used in many
applications.  We partition graphs by balancing the number of edges per-process and minimizing a global edge-cut metric. 
This approach effectively balances per-process workload and helps minimize global communication requirements.


\section{Experimental Setup}
We performed scaling experiments on the AiMOS supercomputer housed at Rensselaer Polytechnic Institute. The system has 268 nodes, each equipped with two IBM Power 9 processors clocked at 3.15~GHz, 4x NVIDIA Tesla V100 GPUs with 16~GB of memory connected via NVLink, 512~GB of RAM, and 1.6~TB Samsung NVMe Flash memory. Inter-node communication uses a Mellanox Infiniband interconnect. We compile with xlC 16.1.1 and use Spectrum MPI with GPU-Direct communication disabled.

The graphs we used to test D1 and D2 are listed in Table~\ref{IAB:tab:graphs}. Most of the graphs are from the SuiteSparse Matrix Collection 
~\cite{IAB:Davis2011UFS}.
The maximum degree $\delta_{max}$ can be considered an upper bound for the number of colors used, as any incomplete, 
connected, and undirected graph can be colored using $\delta_{max}$ colors~\cite{IAB:brooks1941colouring}.
We selected many of the same graphs used by Deveci et al. to allow for direct performance comparisons.
We include many graphs from Partial Differential Equation (PDE) problems because they are representative of graphs used with Automatic Differentiation~\cite{IAB:gebremedhin2020introduction}, which is a target application for graph coloring algorithms.
We also include social network graphs and a web crawl to demonstrate scaling of our methods on irregular real-world datasets.
We preprocessed all graphs to remove multi-edges and self-loops, and 
we used subroutines from HPCGraph~\cite{slota_ipdps2016} for efficient I/O.

We compare our implementation against distributed distance-1 and distance-2 coloring in the Zoltan~\cite{IAB:devine2009getting} package of Trilinos.
Zoltan's implementations are based directly on Bozda{\u{g}} et al.~\cite{IAB:bozdaug2008framework}.
Zoltan's distributed algorithm for distance-2 coloring requires only a single ghost layer, and to reduce conflicts, the boundary vertices are colored in small batches.
For our results, we ran Zoltan and our approaches with four MPI ranks per node on AiMOS, and used the same partitioning method across all of our comparisons.
Our methods D1, D1-2GL, and D2 were run with four GPUs and 
four MPI ranks (one per GPU) per node.
Zoltan uses only MPI parallelism; it does not use GPU or multicore parallelism. 
For consistency, we use four MPI ranks per node with Zoltan, and use the same number of nodes
for experiments with Zoltan and our methods.
We used Zoltan's default coloring parameters; we did not experiment with
options for vertex visit ordering, boundary coloring batch size, etc.


We omit direct comparison to single-node GPU coloring codes such as CuSPARSE~\cite{naumov2015parallel}, as we use subroutines for on-node coloring from Deveci et al.~\cite{IAB:deveci2016parallel}. Deveci et al. have already performed a comprehensive comparison between their coloring methods and those in CuSPARSE, reporting an average speedup of 50\% across a similar set of test instances. As such, we are confident that our on-node GPU coloring is representative of the current state-of-the-art.


\section{Results}


For our experiments, we compare overall performance for D1 and D2 on up to 128 ranks versus Zoltan.
Our performance metrics include execution time, parallel scaling, and number of colors used. 
We do not include the partitioning time for XtraPuLP; we assume target applications will partition and distribute their graphs. 
Each of the results reported represents an average of five runs.

\subsection{Distance-1 Performance}



\begin{figure}[h]
  \centering
  \caption{Performance profiles comparing D1-baseline and D1-recolor-degree on 128 Tesla V100 GPUs with Zoltan's distance-1 coloring on 128 Power9 cores in terms of (a) execution time and (b) number of colors computed for the graphs listed in Table~\ref{IAB:tab:graphs}.} 
  \label{IAB:distance1prof}
  \begin{subfigure}[b]{0.25\textwidth}
    \centering
    \includegraphics[width=\textwidth]{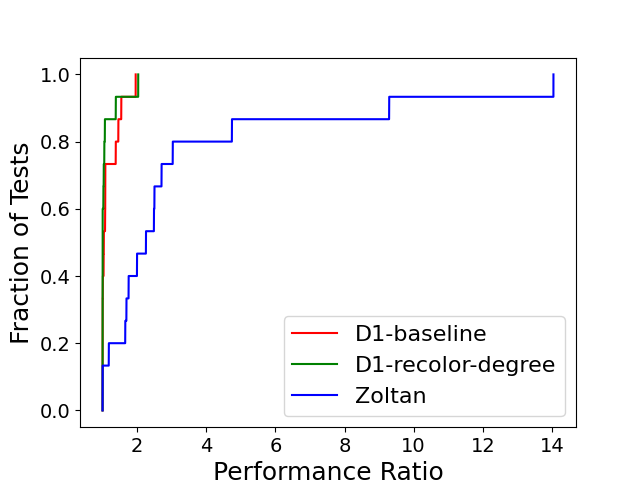}
    \caption{Runtime performance profile}
    \label{IAB:d1runtime}
  \end{subfigure}%
  \begin{subfigure}[b]{0.25\textwidth}
    \centering
    \includegraphics[width=\textwidth]{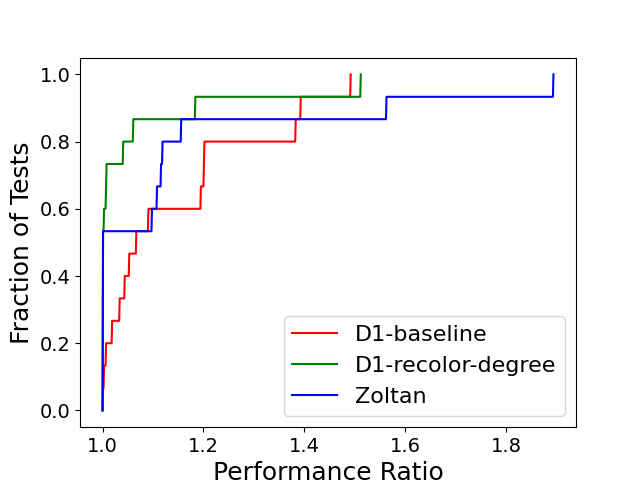}
    \caption{Color performance profile}
    \label{IAB:d1color}
  \end{subfigure}
\end{figure}

We summarize the performance of our algorithms relative to Zoltan 
using the performance profiles in Figure~\ref{IAB:distance1prof}.
Performance profiles plot the proportion of problems an algorithm can solve for a given relative cost.
The relative cost is obtained by dividing each approach's execution time (or colors used) by the best approach's execution time (or colors used) for a given problem.
In these plots, the line that is higher represents the best performing algorithm. 
The further to the right that an algorithm's profile is, the worse it is relative to the best algorithm.
D1-baseline does not consider vertex degree when doing distributed recoloring (e.g., recolorDegrees is false in Algorithm~\ref{IAB:alg:conflictcheck}).
D1-recolor-degree represents our novel approach that recolors distributed conflicts based on vertex degree (e.g., recolorDegrees is true in Algorithm~\ref{IAB:alg:conflictcheck}).

We ran D1-baseline, D1-recolor-degree and Zoltan with 128 MPI ranks to color the 15 SuiteSparse graphs in Table~\ref{IAB:tab:graphs}.
D1-baseline and D1-recolor-degree used MPI plus 128 Tesla V100 GPUs, while Zoltan used MPI on 128 Power9 CPU cores across 32 nodes (four MPI ranks per node).
Some skewed graphs (e.g., hollywood-2009) did not run on 128 ranks on Zoltan or D1-baseline; in those cases we use the largest run that completed for both approaches.
Figure~\ref{IAB:d1runtime} shows that D1-recolor-degree outperforms both Zoltan and D1-baselines in terms of execution time in these experiments.
D1-baseline and D1-recolor-degree are very similar in terms of runtime performance, but D1-recolor-degree is the fastest approach for 60\% of the graphs,
D1-baseline is fastest on 26\%, and Zoltan is fastest on 13\%.
Zoltan is faster than our approaches on two of the smallest graphs, Audikw\_1, and ldoor.
D1-baseline is faster than D1-recolor-degrees on four graphs which are more varied in application and structure: Bump\_2911, com-Friendster, rgg\_n\_2\_24\_s0, and twitter7.
There are four graphs for which D1-baseline and D1-recolor-degrees runtime performance differ substantially: Audikw\_1 (D1-baseline is 32\% faster), ldoor (D1-recolor-degrees is 42\% faster), mycielskian19 (D1-recolor-degrees is 45\% faster), and mycielskian20 (D1-recolor-degrees is 38\% faster).
D1-baseline has at most a 10x speedup over Zoltan (with the mycielskian20 graph) and at worst an 1.95x slowdown relative to Zoltan (with ldoor), while
D1-recolor-degrees achieves at most a 14x speedup over Zoltan (on mycielskian20), and at worst a 2x slowdown (on Audikw\_1).

Figure~\ref{IAB:d1color} shows that Zoltan outperforms D1-baseline in terms of color usage, but D1-recolor-degree is much more competitive.
Both Zoltan and D1-recolor-degree use the fewest colors in 53\% of experiments; Zoltan and D1-recolor-degree tie on a single graph.
D1-baseline uses the fewest number of colors on a single graph, for which it ties D1-recolor-degree.
D1-recolor-degree uses more colors than D1-baseline for two graphs (indochina-2004 and twitter7); in both graph, D1-baseline uses roughly 1\% fewer colors.
On average, D1-recolor-degree  uses 8.9\% fewer colors than D1-baseline, 
and in the best case, it reduces color usage  39\% relative to D1-baseline (mycielskian19).
On average, D1-recolor-degree uses 4\% fewer colors than Zoltan.
In the worst case, D1-recolor-degree uses 51\% more colors than Zoltan (twitter7); in the best case,
D1-recolor-degree uses 53\% fewer colors than Zoltan (mycielskian20).

Because the performance of D1-recolor-degree is generally better than that of D1-baseline, 
all further distance-1 coloring results use D1-recolor-degree, and we refer to D1-recolor-degree as D1 going forward.

\subsection{Distance-1 Strong Scaling}

\begin{figure}[h]
  \centering
  \caption{Zoltan and D1 strong scaling on select (a) PDE and (b) Social Network graphs.}
  \label{IAB:realstrong}
  \begin{subfigure}[b]{0.25\textwidth}
    \centering
    \includegraphics[scale=0.5]{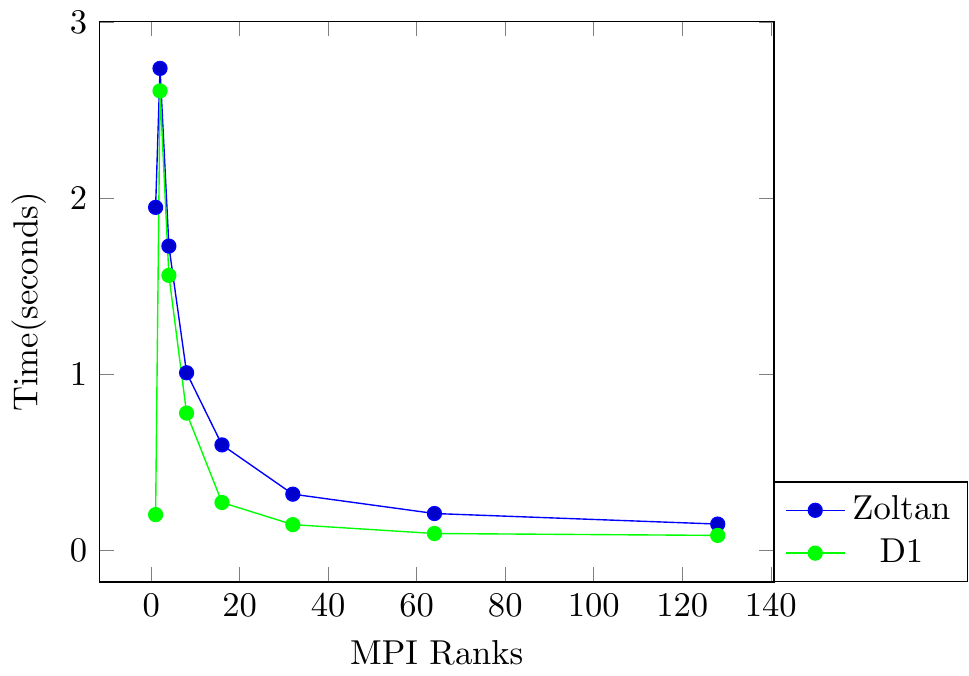}
    \caption{Queen\_4147}
    \label{IAB:queenhybridzoltan}
  \end{subfigure}%
  \begin{subfigure}[b]{0.22\textwidth}
    \centering
    \includegraphics[scale=0.5]{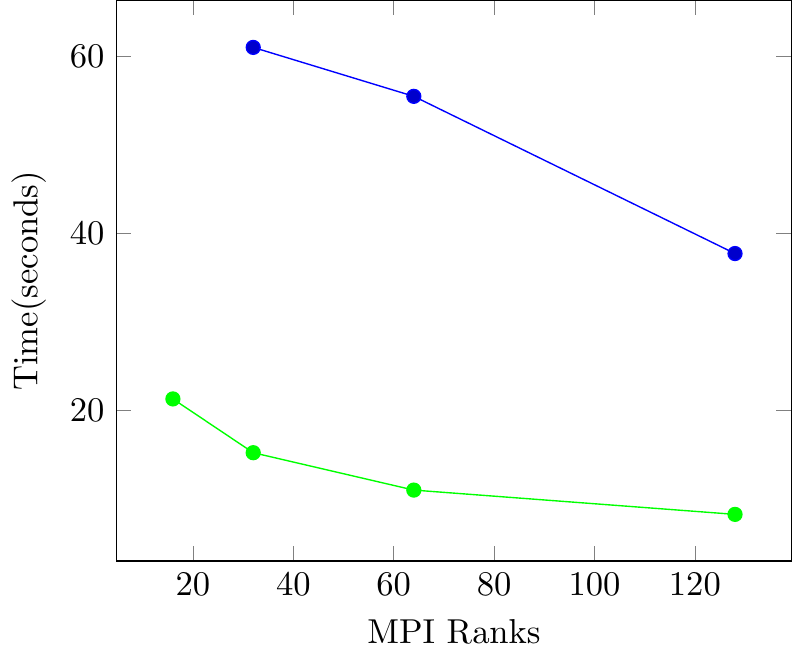}
    \caption{com-Friendster}
    \label{IAB:friendsterhybridzoltan}
  \end{subfigure}
\end{figure}

Figure~\ref{IAB:realstrong} shows strong scaling times for Queen\_4147 and com-Friendster.
These graphs are selected for presentation because they are the largest graphs of their respective problem domains.
Data points that are absent were the result of out-of-memory issues or execution times (including partitioning) 
that were longer than our single job allocation limits.
D1 scales better on the com-Friendster graph than on Queen\_4147, as 
the GPUs can be more fully utilized with the much larger com-Friendster graph. 
For Queen\_4147, D1 on 128 GPUs shows a speedup of around 2.38x over a single GPU.
D1 uses 12\% fewer colors than Zoltan in the 128 rank run on Queen\_4147, as well as running 1.75x faster than Zoltan on that graph.
For com-Friendster, D1 is roughly 4.6x faster than Zoltan in the 128 rank run, and only uses 0.6\% more colors than Zoltan.

For graph processing in general, it is often difficult to demonstrate good strong scaling relative to single node runs. From the Graph500.org benchmark (June 2020 BFS results)~\cite{graph500}, the relative per-node performance difference in the metric of ``edges processed per second'' between the fastest multi-node results and fastest single node results are well over 100x.
For coloring on GPUs, graphs that can fit into a single GPU do not provide sufficient work parallelism for large numbers of GPUs, and multi-GPU execution incurs communication overheads and additional required rounds for speculative coloring.
However, on roughly half of the graphs that fit on a single GPU, D1 with 128 GPUs achieves an average speedup of 1.9x over a single GPU.
D1 achieves a maximum speedup of 2.43x on the mycielskian20 graph.
For the other half of the graphs, D1 does not show a speedup over a single GPU.
On small or highly skewed graphs that fit on a single GPU, speedup is limited, due to the communication overheads and work imbalances that result from distribution even with relatively good partitioning.
Distributed coloring is valuable even for these small problems, however, as parallel applications 
using coloring typically have distributed data that would be expensive to gather into one GPU for single-GPU coloring.

On average over all the graphs, D1 uses 38\% more colors than the single GPU run, while Zoltan uses 53.6\% more colors than the single GPU run.
Such large color usage increases are mostly due to the Mycielskian19 and Mycielskian20 graphs.
These graphs were generated to have known minimum number of colors (chromatic numbers) of 19 and 20 respectively, and our single GPU runs use 19 and 21 colors to color those graphs.
Both D1 and Zoltan have trouble coloring these graphs in distributed memory, but our D1 implementation colors these graphs in fewer colors than Zoltan.
Without these two outliers, the average color increase from the single GPU run is only 2.23\% for D1, and Zoltan decreases color usage by 0.1\% on average.
Zoltan's higher coloring quality is due to its inherently lower concurrency.

\begin{figure}[h]
  \centering
  \caption{D1 communication time (Comm) and computation time (Comp) from 1 to 128 GPUs.}
  \label{IAB:strongbreakdown}
  \begin{subfigure}[b]{0.25\textwidth}
    \centering
    \includegraphics[scale=0.5]{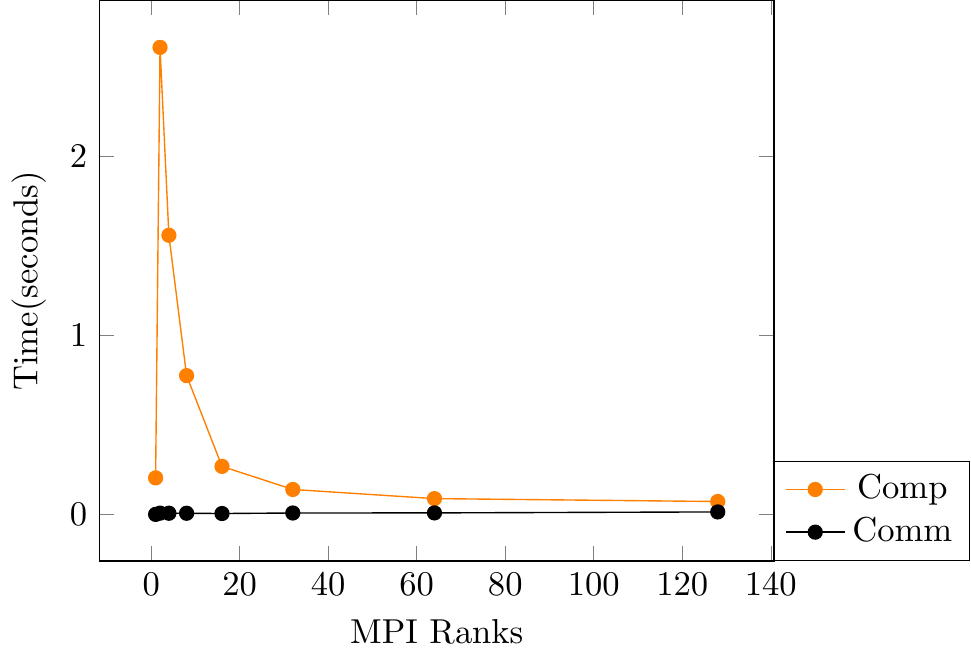}
    \caption{Queen\_4147}
    \label{IAB:queenbreakdown}
  \end{subfigure}%
  \begin{subfigure}[b]{0.23\textwidth}
    \centering
    \includegraphics[scale=0.5]{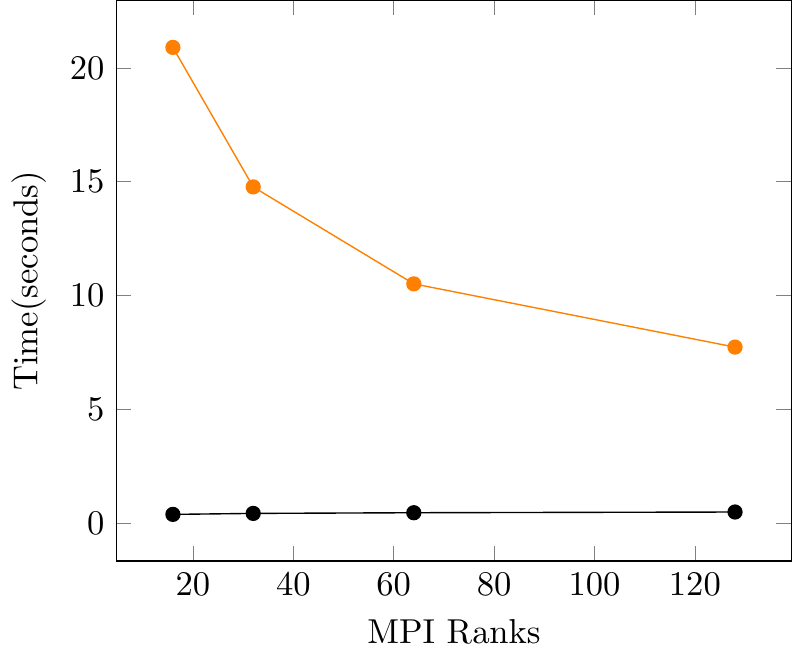}
    \caption{com-Friendster}
    \label{IAB:friendsterbreakdown}
  \end{subfigure}
  \hspace*{\fill}
\end{figure}

Figure~\ref{IAB:strongbreakdown} shows the total communication and computation time associated with each run.
For both graphs, the dominant scaling factor is computation. 
Specifically, the computational overhead associated with recoloring vertices in distributed memory is the dominant scaling factor.
However, strong scaling is good on both graphs,
despite the fact that adding more ranks to a problem also increases the number of vertices that need to be recolored.
Figure~\ref{IAB:friendsterbreakdown} shows that D1 scales to more ranks on com-Friendster, primarily because of the graph's larger size.

\subsection{Distance-1 Weak Scaling}

The greatest benefit of our approach is its ability to efficiently process massive-scale graphs. 
We demonstrate this benefit with a weak-scaling study conducted with uniform 3D hexahedral meshes.
The meshes were partitioned with block partitioning along a single axis, resulting in the mesh being distributed in ``slabs.''
Larger meshes were generated by doubling the number of elements in a single dimension to keep the per-process communication and computational workload constant.
Each distinct per-process workload increases the boundary by a factor of two, which correspondingly increases communication and recoloring overhead for distributed runs.
We run with up to 100 million vertices per GPU, yielding a graph of 12.8 billion vertices and 76.7 billion edges in our largest tests; \textbf{this graph was colored in less than two seconds}.

\begin{figure}[h]
  \centering
  \caption{Weak scaling of D1 on 3D mesh graphs. Tests use 12.5, 25, 50, and 100 million vertices per GPU.}
  \includegraphics[scale=0.6]{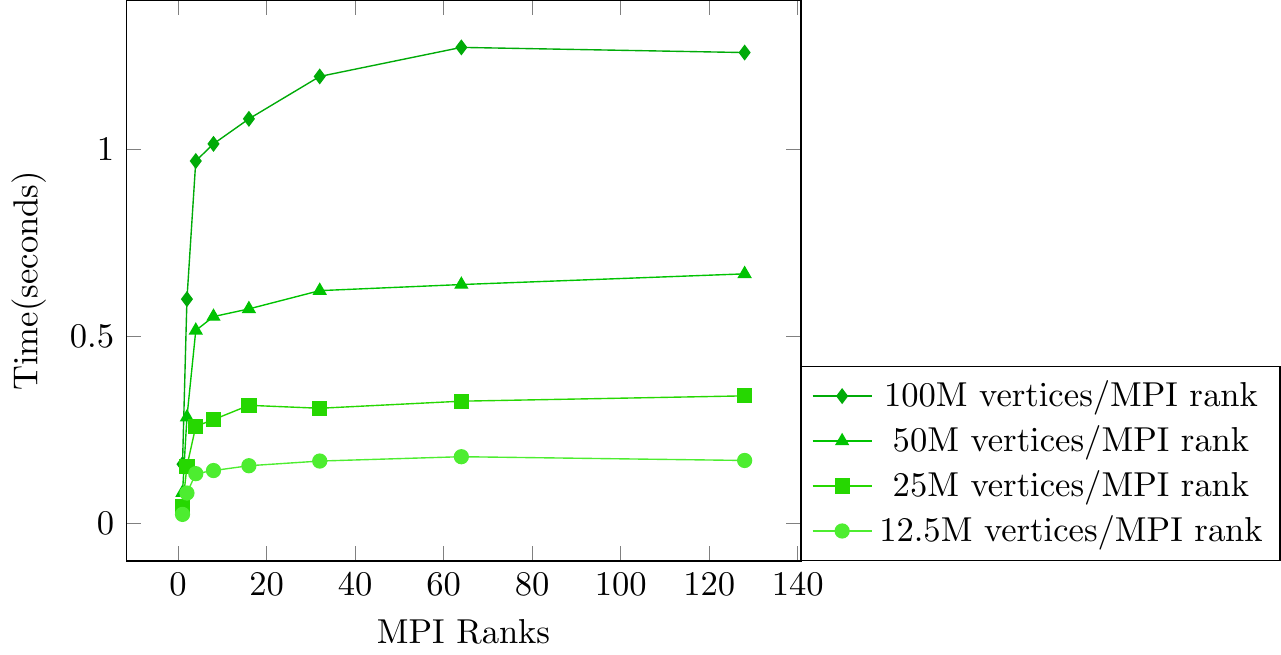}
  \label{IAB:meshhybridweak}
\end{figure}

Figure~\ref{IAB:meshhybridweak} shows that the single rank runs for each workload are similar, 
indicating that communication and recoloring overhead are the dominant scaling factors for this study.
In increasing the boundary size by a factor of two, we do not necessarily increase the number of distributed conflicts by two,
especially in such a regular graph.
The smaller workloads all have similar and relatively small recoloring workloads, which is why they show more consistent weak scaling than 
the 100 Million vertex per rank experiment. 
That particular experiment does substantially more recoloring than the others, resulting in its increase in runtime as the number of ranks increases.
We have found that for extremely regular meshes like these, the number of vertices on process boundaries impacts the recoloring workloads for D1.

\subsection{D1-2GL Performance}

\begin{figure}[h]
  \centering
  \caption{Number of communication rounds for D1-baseline and D1-2GL on Queen\_4147 from 2 to 128 ranks.}
  \includegraphics[scale=0.6]{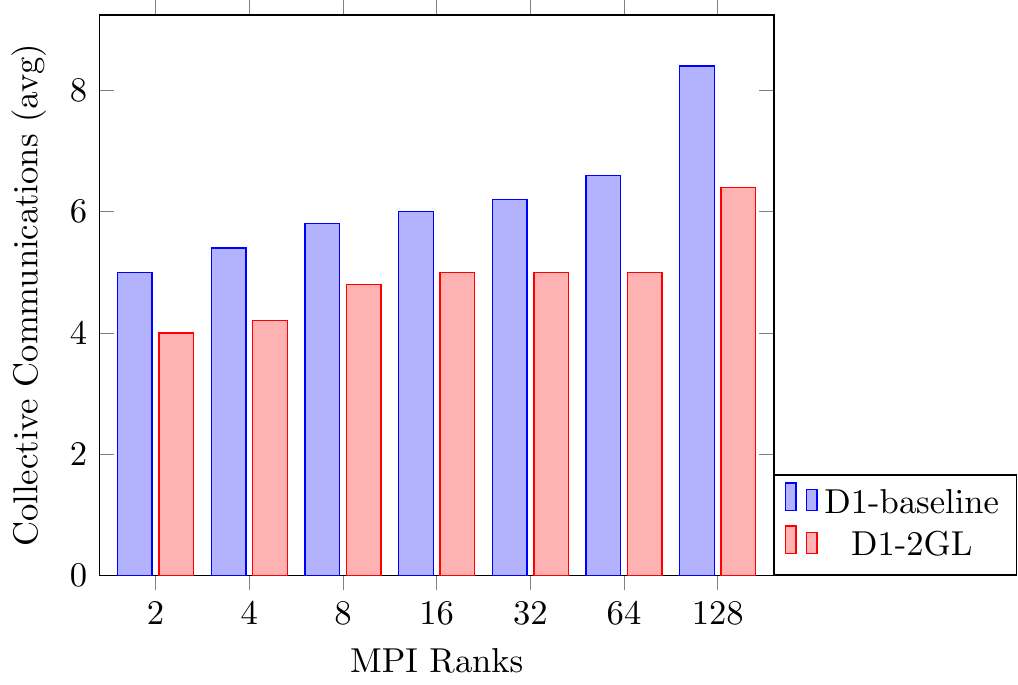}
  \label{fig:2GLrounds}
\end{figure}

In general, D1-2GL reduces the number of collective communications used in the distributed distance-1 coloring. Figure~\ref{fig:2GLrounds} compares the number of communication rounds for D1-baseline and D1-2GL on the Queen\_4157 input for 2 to 128 MPI ranks, averaged over five runs. 
With 128 ranks on this graph, D1-2GL method reduces the number of rounds by 25\% on average, giving 
speedup of 1.18x. 
D1-2GL provides speedups over D1-baseline with the smaller graphs: 1.17x with Audikw\_1 and 1.2x with ldoor.
Unfortunately, due to the increased cost of each communication round, D1-2GL does not generally achieve a total execution time speedup over D1-baseline on AiMOS. 
Additionally, second ghost layer vertices may be recolored if they are boundary vertices on another processor; this occurs often in dense inputs and incurs further recoloring rounds.
However, in distributed systems with much higher latency costs, D1-2GL could be beneficial.


\subsection{Distance-2 Performance}

\begin{figure}[h]
  \centering
  \caption{Performance profiles comparing D2 on 128 Tesla V100 GPUs with Zoltan's distance-2 coloring on 128 Power9 cores in terms of (a) execution time and (b) number of colors computed for a subset of graphs listed in Table~\ref{IAB:tab:graphs}.} 
  \label{IAB:distance2prof}
  \begin{subfigure}[b]{0.25\textwidth}
    \centering
    \includegraphics[width=\textwidth]{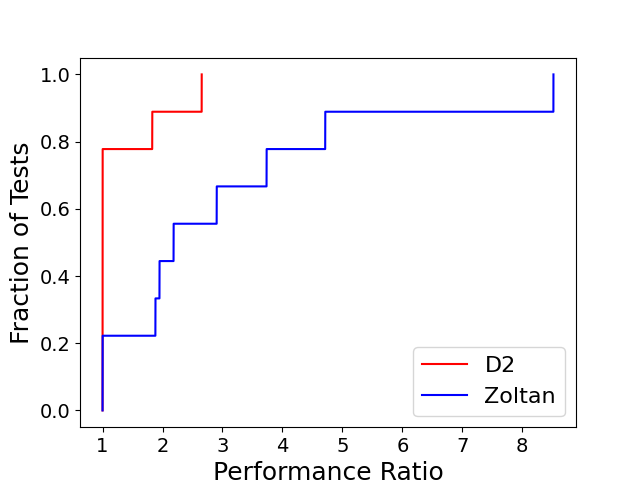}
    \caption{Runtime performance profile}
    \label{IAB:d2runtime}
  \end{subfigure}%
  \begin{subfigure}[b]{0.25\textwidth}
    \centering
    \includegraphics[width=\textwidth]{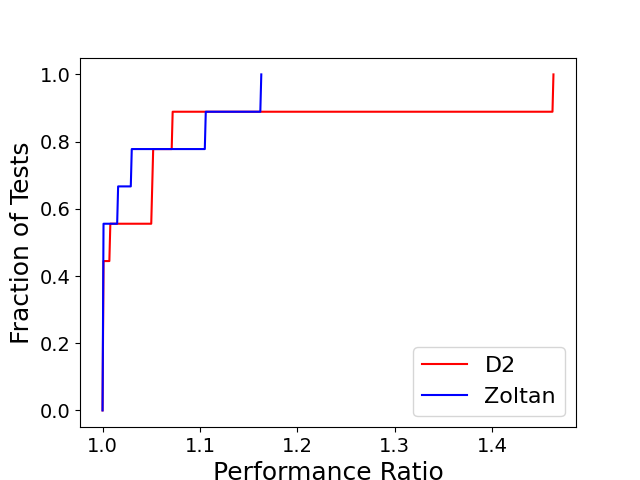}
    \caption{Color performance profile}
    \label{IAB:d2color}
  \end{subfigure}
\end{figure}

We compare our D2 method to Zoltan's distance-2 coloring using eight graphs from Table~\ref{IAB:tab:graphs}:  Bump\_2911, Queen\_4147, hollywood-2009, europe\_osm, rgg\_n\_2\_24\_s0, ldoor, Audikw\_1, and soc-LiveJournal1.
We use the same experimental setup as with the distance-1 performance comparison.
Figure~\ref{IAB:d2runtime} shows that D2 compares well against Zoltan in terms of execution time, with D2 outperforming Zoltan on all but two graphs.
In the best case, we see an 8.5x speedup over Zoltan on the Queen\_4147 graph.

Figure~\ref{IAB:d2color} shows that D2 has similar color usage as Zoltan.
D2 and Zoltan each produce fewer colors in half of the experiments.
In all but one case in which Zoltan uses fewer colors, D2 uses no more than 10\% more colors.
Interestingly, the number of colors used by D2 on the soc-LiveJournal1 graph is unchanged with one and 128 GPUs.
Zoltan outperforms D2 with respect to runtime on skewed graphs because Zoltan has distance-2 optimizations which reduce communication overhead
and minimize the chance for distributed conflicts.

\subsection{Distance-2 Strong Scaling}

\begin{figure}[h]
  \centering
  \caption{D2 and Zoltan strong scaling for distance-2 coloring.}
  \label{IAB:distance2strong}
  \begin{subfigure}[b]{0.25\textwidth}
    \centering
    \includegraphics[scale=0.5]{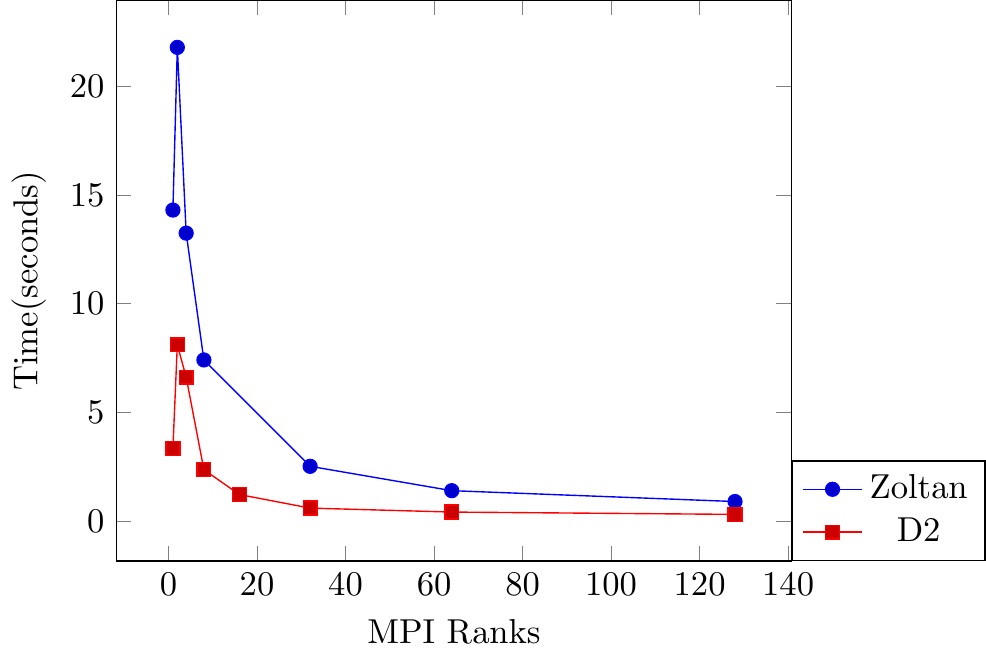}
    \caption{Bump\_2911}
    \label{IAB:Bumpstrong}
  \end{subfigure}%
  \begin{subfigure}[b]{0.22\textwidth}
    \centering
    \includegraphics[scale=0.5]{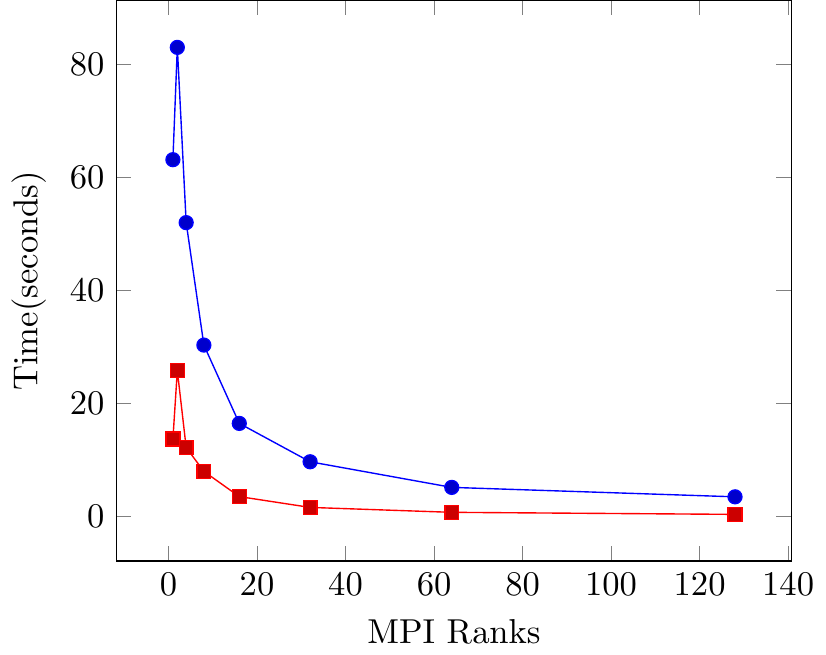}
    \caption{Queen\_4147}
    \label{IAB:queenstrong}
  \end{subfigure}
\end{figure}

Figures~\ref{IAB:Bumpstrong} and~\ref{IAB:queenstrong} show the strong scaling behavior of D2 and Zoltan on Bump\_2911 and Queen\_4147.
Bump\_2911 shows that D2 scales better initially than Zoltan, and with 128 ranks, D2 is 2.9x faster than Zoltan, using 0.7\% more colors.
Queen\_4147 shows better scaling for D2 as well; with 128 ranks, D2 is 8.5x faster than Zoltan and uses 10\% fewer colors.


On average over the eight graphs, D2 exhibits 4.29x speedup on 128 GPUs over a single GPU, and uses 7.5\% more colors than single GPU runs.
Speedup is greater with D2 than D1 because distance-2 coloring is more computationally intensive, and thus has a larger work-to-overhead ratio.

\begin{figure}[h]
  \centering
  \caption{D2 communication time (comm) and computation time (comp) from 1 to 128 GPUs.}
  \label{IAB:distance2breakdown}
  \begin{subfigure}[b]{0.25\textwidth}
    \centering
    \includegraphics[scale=0.5]{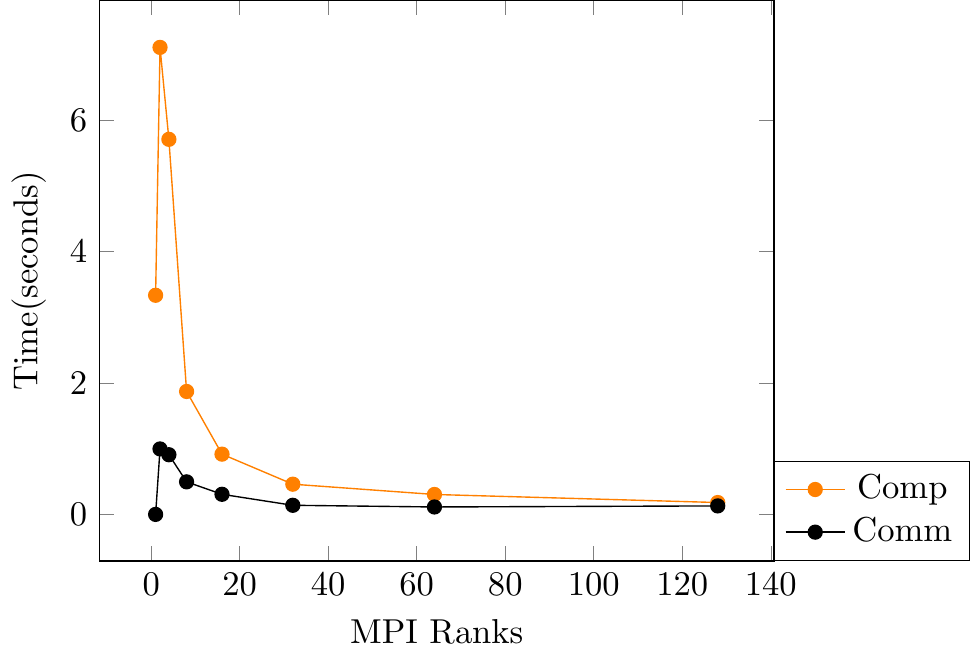}
    \caption{Bump\_2911}
    \label{IAB:bumpbreakdown}
  \end{subfigure}%
  \begin{subfigure}[b]{0.22\textwidth}
    \centering
    \includegraphics[scale=0.5]{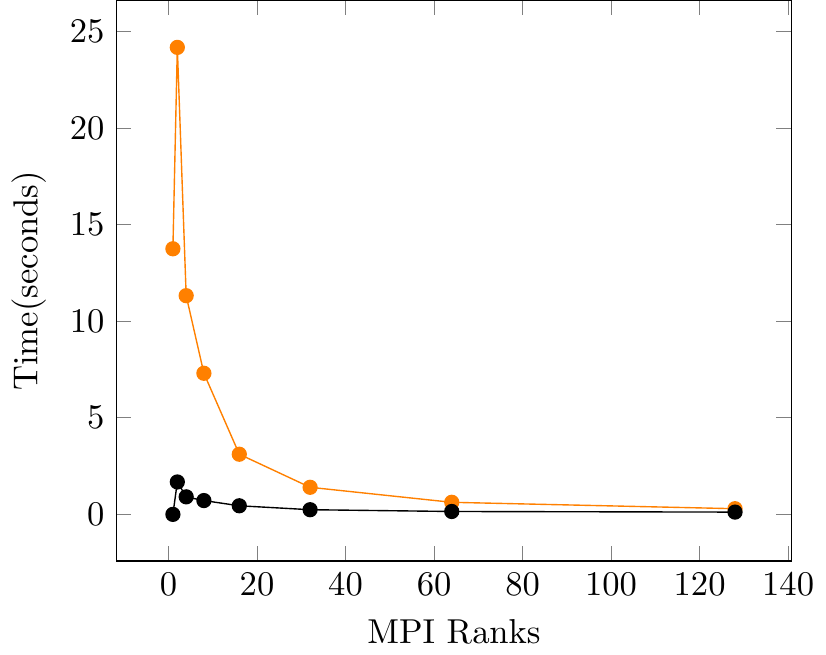}
    \caption{Queen\_4147}
    \label{IAB:d2queenbreakdown}
  \end{subfigure}
\end{figure}

Figures~\ref{IAB:bumpbreakdown} and~\ref{IAB:d2queenbreakdown} show the communication and computation breakdown of D2 on Bump\_2911 and Queen\_4147.
Bump\_2911 shows computation and communication scaling for up to 128 ranks, while color usage increases by only 0.6\%.
In general, the relative increase in color usage from a single rank for distance-2 coloring is less than for distance-1 coloring. The number of colors used for distance-2 coloring is greater than for distance-1; therefore, a similar absolute increase in color count results in a lower proportional increase. 

\subsection{Distance-2 Weak Scaling}

\begin{figure}[h]
  \centering
  \caption{Distance-2 weak scaling of D2 on 3D mesh graphs.}
  \includegraphics[scale=0.6]{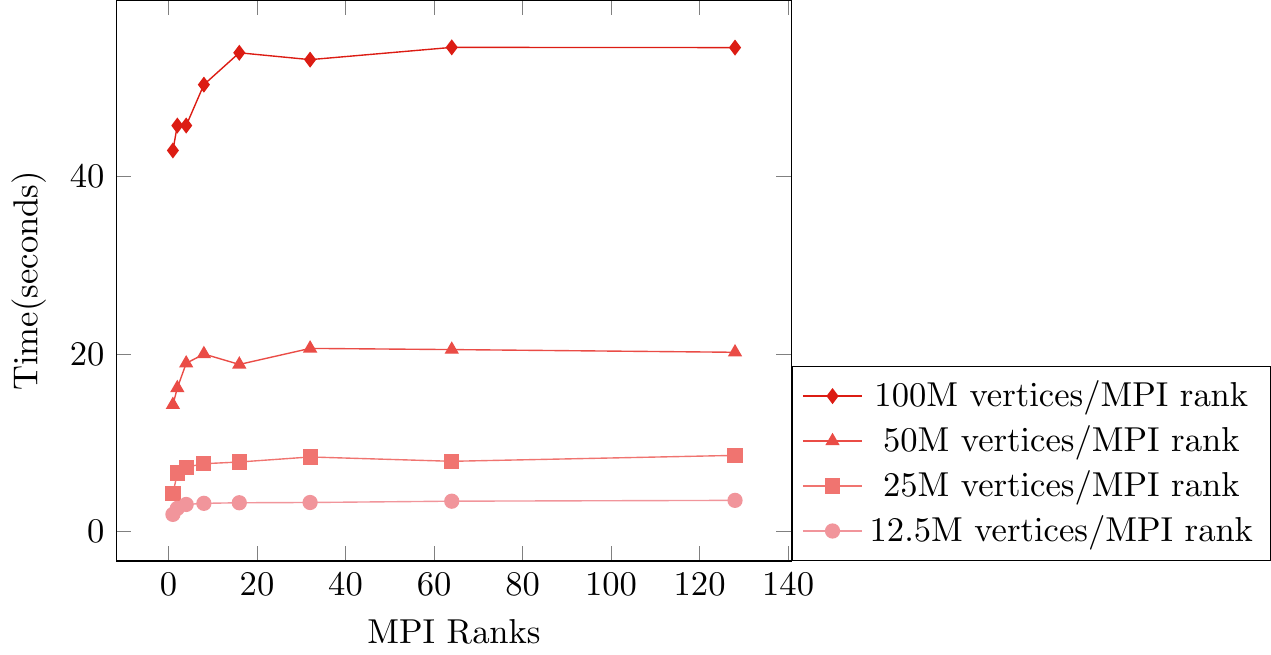}
  \label{IAB:meshd2weak}
\end{figure}

Figure~\ref{IAB:meshd2weak} demonstrates the weak scaling behavior for D2.
The same hexahedral mesh graphs were used as in the D1 weak scaling experiments.
In general, D2 has fairly consistent weak scaling.
The runtimes across workloads with a single rank increase by more than factor of two because the number of edges in each mesh increases by more than a factor of two, and the complexity of the distance-2 coloring algorithm for local colorings depends on the number of edges.
Weak scaling to large process counts is good for all workloads.

\subsection{Partial Distance-2 Strong Scaling}

\begin{table}[!t]
  \small
  \centering
  \caption{Summary of the graphs used for PD2 tests. Statistics are for the bipartite representation of the graph (Section ~\ref{IAB:method:PD2}).  $\delta_{avg}$ is average degree and $\delta_{max}$ is maximum degree. Numeric values listed are after preprocessing to remove multi-edges and self-loops. k = thousand, M = million}
  \begin{tabular}{|r|r|r|r|r|r|r|}
    \hline
    Graph	& Class	&	\#Vtx	& \#Edges	& $\delta_{avg}$ & $\delta_{max}$ \\
    \hline
    Hamrle3	&Circuit Sim. &2.9 M	&5.5 M		& 3.5		 & 18		  \\
    patents	&Patent Citations &7.5 M	&14.9 M		& 1.9		 & 1k		  \\
    \hline
  \end{tabular}\\
  \label{IAB:tab:pd2graphs}
\end{table}

Table~\ref{IAB:tab:pd2graphs} shows the graphs that we used to compare our PD2 implementation against Zoltan.
Partial distance-2 coloring is typically used on non-symmetric and bipartite graphs; the graphs in Table ~\ref{IAB:tab:pd2graphs} are representative of application use cases. 
We report metrics reported for the bipartite representation of the graph (as described in Section~\ref{IAB:method:PD2}).
Partial distance-2 colorings typically are needed for only a subset of the vertices in a graph, but 
our PD2 implementation colors all vertices in the graph. We compare to Zoltan, which colors only vertices that would be colored in typical partial distance-2 coloring. 
Thus, in general, PD2 is coloring roughly twice as many vertices as Zoltan.

\begin{figure}[h]
  \centering
  \caption{PD2 strong scaling for partial distance-2 coloring.}
  \label{IAB:partialdistance2strong}
  \begin{subfigure}[b]{0.25\textwidth}
    \centering
    \includegraphics[scale=0.5]{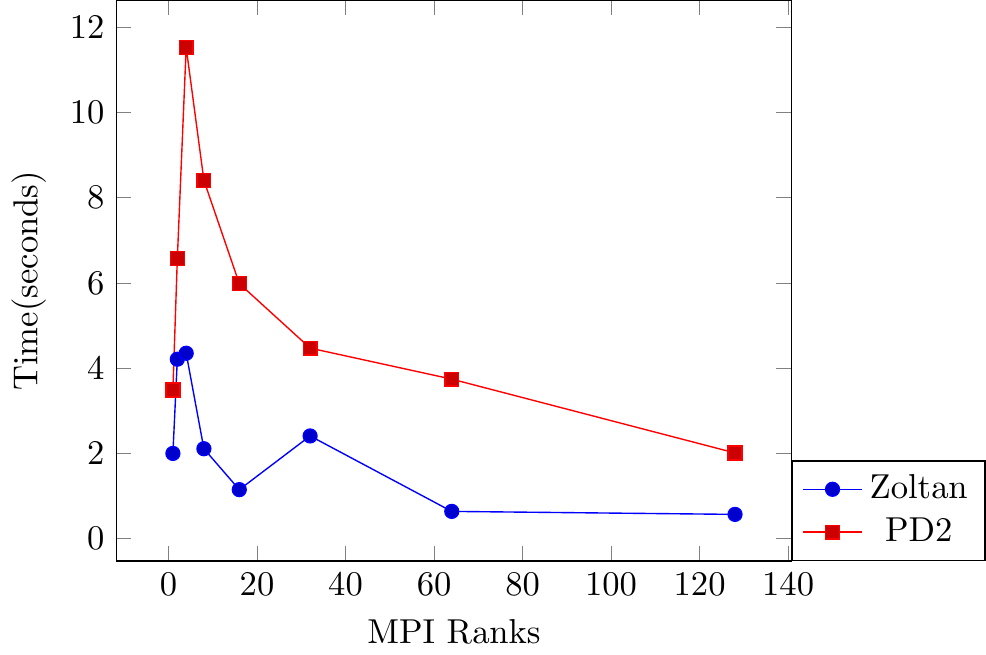}
    \caption{patents}
    \label{IAB:patentsstrong}
  \end{subfigure}%
  \begin{subfigure}[b]{0.22\textwidth}
    \centering
    \includegraphics[scale=0.5]{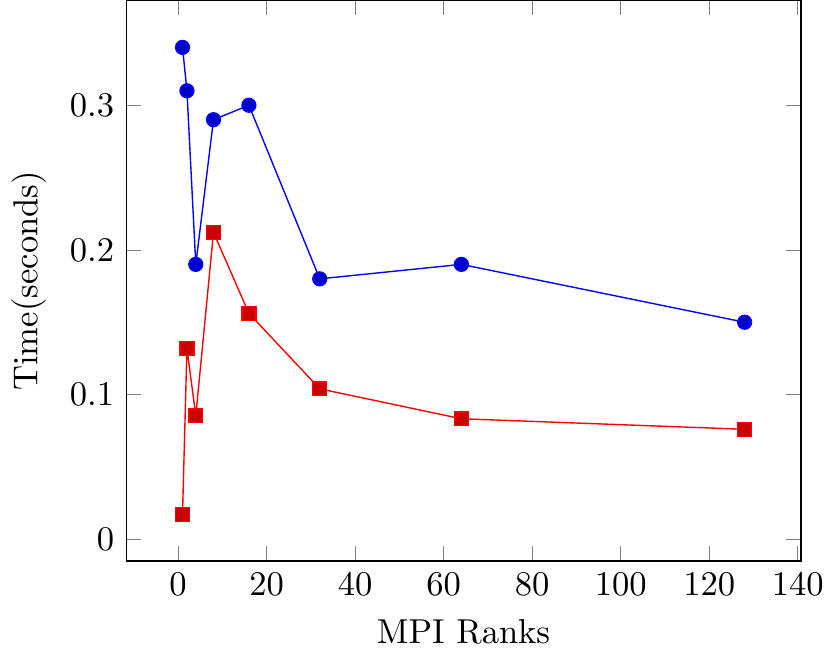}
    \caption{Hamrle3}
    \label{IAB:hamrle3strong}
  \end{subfigure}
\end{figure}

Figure~\ref{IAB:partialdistance2strong} shows the strong scaling behavior of PD2.
The experiment with Hamrle3 on four ranks benefits from a particularly good partition that results in less recoloring for both PD2 and Zoltan relative to other process configurations.
For the patents graph, D2 has a particularly heavy recoloring workload for four ranks, resulting in a large increase in runtime from two to four ranks.
Even though PD2 is coloring more vertices than Zoltan in these tests, PD2 achieves roughly 2x speedup on 128 ranks with Hamrle3.
With patents, Zoltan is faster than PD2;  this result can be attributed partially to Zoltan's optimized recoloring scheme that reduces the number of conflicts introduced while recoloring distributed conflicts.
PD2 achieves a 1.73x speedup over a single GPU with the patents graph, while it did not show any speedup from a single GPU with Hamrle3.
Figure~\ref{IAB:patentsstrong} shows that, with the patents graph, Zoltan is faster on one core than a single GPU.
This speedup is attributed to Zoltan's coloring fewer vertices than PD2;
when Zoltan colors the same number of vertices as PD2, their single rank runtimes are equal.
Investigating the cause of this result is a subject for future research.

For these two graphs, PD2 uses a very similar number of colors as Zoltan. 
PD2 uses at most 10\% more colors in the distributed runs.
This difference is typically only one to five colors more than Zoltan.

\begin{figure}[h]
  \centering
  \caption{PD2 communication time (comm) and computation time (comp) from 1 to 128 GPUs}
  \label{IAB:partialdistance2breakdown}
  \begin{subfigure}[b]{0.25\textwidth}
    \centering
    \includegraphics[scale=0.5]{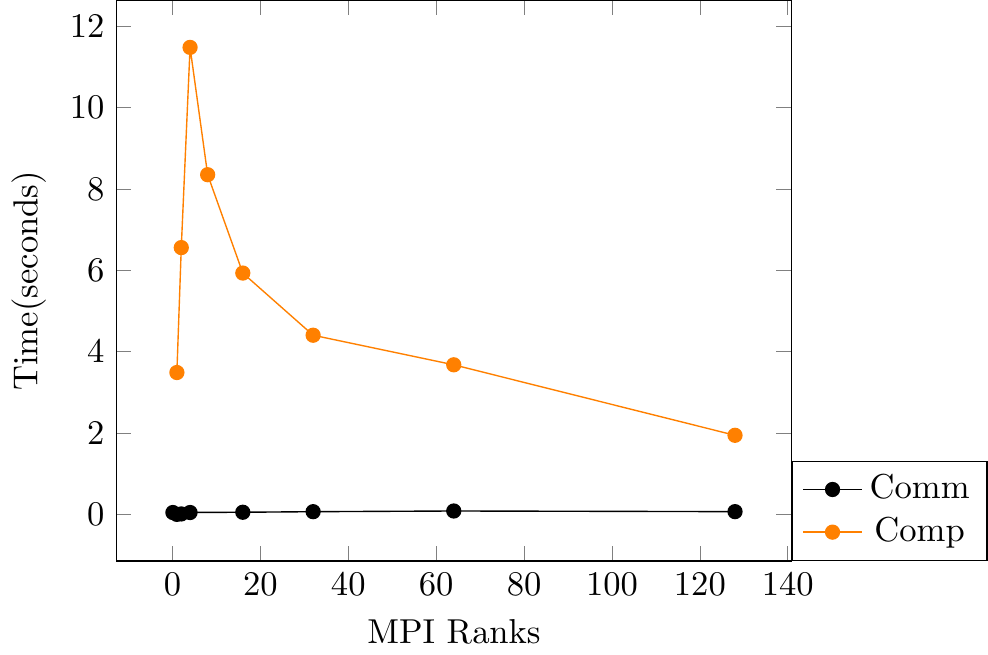}
    \caption{patents}
    \label{IAB:patentsbreakdown}
  \end{subfigure}%
  \begin{subfigure}[b]{0.22\textwidth}
    \centering
    \includegraphics[scale=0.5]{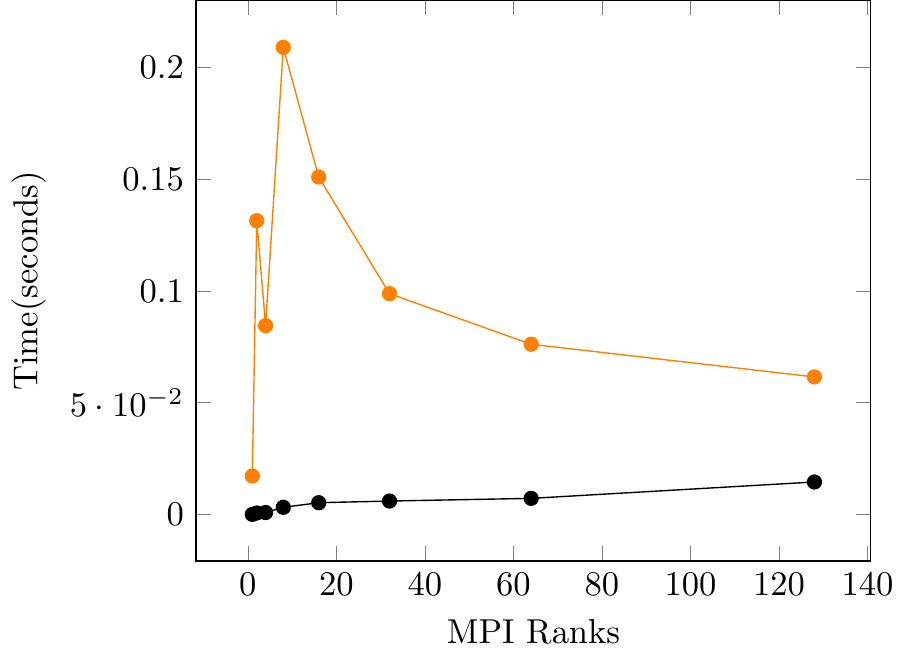}
    \caption{Hamrle3}
    \label{IAB:hamrle3breakdown}
  \end{subfigure}
\end{figure}

Figure~\ref{IAB:partialdistance2breakdown} shows that computation is the main factor in the scaling behavior of PD2.
In distributed runs, the largest factor of the runtime is the computation overhead involved in recoloring distributed conflicts.
Figure~\ref{IAB:hamrle3breakdown} shows an unexpected decrease in computation for the Hamrle3 graph for four ranks, 
which is due to a decrease in the recoloring workload. 
PD2's recoloring workload is approximately 25,000 vertices per rank in most experiments, but the four-rank experiment has a recoloring workload of 9,000 vertices per rank.
Figure~\ref{IAB:patentsbreakdown} shows that the four-rank PD2 run has a much longer computation time than expected; this is due to the total distributed recoloring workload increasing by a factor of six.
Additionally, the 64-rank run with the patents graph shows slightly less computational scaling than expected, due to an increase in recoloring rounds.
Increasing recoloring rounds serializes recoloring computation and incurs more rounds of communication, resulting in a runtime increase.
Optimizing recoloring to reduce subsequent conflicts and reduce the number of recoloring rounds necessary in D2 and PD2 are subjects for future research.

\section{Future work}

We have presented new multi-GPU distributed memory implementations of distance-1, distance-2 and
partial distance-2 graph coloring.  These methods enable parallel graph coloring for graphs too large
to fit into a single GPUs memory; weak-scaling results demonstrate coloring of a graph with
12.8 billion vertices and 76.7 billion edges in less than two seconds.  
We introduced a new recoloring heuristic based on vertex degrees 
that reduces the amount of recoloring needed in parallel coloring methods.
We showed that our approaches are scalable to 128 GPUs and produce colorings with quality
similar to or better than Zoltan's distributed memory coloring algorithms.

Because our coloring algorithms use the Kokkos and KokkosKernels library for 
on-node performance portability, our MPI+X methods can also run on distributed-memory 
computers with multicore (CPU-based) nodes.
In this work, we focused on GPU architectures; exploring multicore performance 
is future work.

We are currently integrating this code into the Zoltan2 package of Trilinos.
Our goal is to deliver a complete suite of MPI+X algorithms for distance-1, distance-2, and partial distance-2 coloring in Zoltan2. 
We will modify PD2 to allow it to color only vertices of interest to the application as Zoltan does.


We also will investigate further optimizations to increase performance.
There are optimizations present in Zoltan's implementation that are not directly applicable to our implementation, but can inform optimizations that
reduce the overall recoloring workload and minimize communication.
These changes could increase performance for D1, D2, and PD2, 
as well as make D2 and PD2 more scalable on skewed graphs.

\section{Acknowledgments}


We thank the Center of Computational Innovations at RPI for maintaining the equipment used in this research, including the AiMOS supercomputer supported by the National Science Foundation under Grant No. 1828083.
This research was also supported by the Exascale Computing Project (17-SC-20-SC), a collaborative effort of the U.S. Department of Energy Office of Science and the National Nuclear Security Administration. Sandia National Laboratories is a multimission laboratory managed and operated by National Technology and Engineering Solutions of Sandia, LLC., a wholly owned subsidiary of Honeywell International, Inc., for the U.S. Department of Energy's National Nuclear Security Administration under contract DE-NA-0003525.


\clearpage
\section*{References}
\bibliographystyle{elsarticle-num}
\bibliography{bib}

\end{document}